\documentclass[10pt,onecolumn]{article} 
\usepackage[utf8]{inputenc}
\usepackage{style_arxiv}
\usepackage{amsmath}
\usepackage{amsfonts}
\usepackage{amssymb}
\usepackage{url}
\usepackage[symbol*]{footmisc}
\setfnsymbol{wiley}
\renewcommand{\vec}[1]{\mathbf{#1}}
\ifx\pdfoutput\undefined
	\usepackage[dvips]{graphicx}
\else
	\usepackage[pdftex]{graphicx}
	\usepackage{epstopdf,subfigure}
\fi
\graphicspath{./}
\newcounter{subfig}
\begin{document}

\title{Modeling the potential interaction energy of two atoms with hydrogen-like wave functions}

\author{Vladimir.~P.~Koshcheev$^{1,}$\email{koshcheev1@yandex.ru}~, Dmitry.~A.~Morgun$^{2,}$\email{morgun\_da@office.niisi.tech}~~, Yuriy.~N.~Shtanov$^{3,}$\email{yuran1987@mail.ru}}

\affiliation{\small{\textit{$^1$Moscow Aviation Institute (National Research University),}}\\ 
\small{\textit{Strela Branch, Moscow oblast, Zhukovskii, 140180 Russia}}\\
\small{\textit{$^2$Surgut Branch of Federal State Institution "Scientific Research}} \\
\small{\textit{Institute for System Analysis of the Russian Academy of Sciences", Surgut, 628408, Russia}}\\
\small{\textit{$^3$Tyumen Industrial University, Surgut Branch, Surgut, 628404 Russia}}}

\preparedfor{}

\maketitle
\thispagestyle{empty}

\begin{abstract}
The Fourier component of the potential energy of interaction of an atom with an atom is represented as a polynomial of the fourth degree from the atomic form factor. A numerical calculation was performed for the atomic form factor in the approximation of hydrogen-like wave functions. It is shown that taking the Pauli principle into account leads to two local potential wells, which are separated by a potential barrier. It was shown that this model gives satisfactory agreement with the experiment in the depth of the potential well, but its position and width differ from the results of the experiment with a diatomic beryllium molecule (dimer). It is shown that the further approach of two beryllium atoms leads to a new potential minimum, which could not be found in the literature.
\end{abstract}
\par 
\\
\\
\\
\par The density functional method continues to be the main method for calculating the interaction between atoms (see, for example, \cite{CHEN2020100731}) despite criticism of the method \cite{Medvedev49,Sarry2012}. The search for other methods for calculating the interaction between atoms remains relevant. In \cite{Koshcheev2018}, an alternative solution to this problem was proposed, in which taking the Pauli principle into account leads to a potential barrier and an additional attraction region of two atoms. Further development of the approach \cite{Koshcheev2018} is presented in this paper.
\par Consider the potential energy of interaction of an atom with a charge $Z_1 e$ with an atom whose charge $Z_2 e$
\begin{equation}\label{eq1} 
U=\frac{Z_{1} Z_{2} e^{2} }{\left|\vec{r}_1-\vec{r}_2\right|}+\sum_{j_1=1}^{Z_1}\sum_{j_2=1}^{Z_2}\frac{e^2}{\left|\vec{r}_{1j_1}-\vec{r}_{2j_2} \right|}-\sum _{j_2=1}^{Z_2}\frac{Z_1 e^2}{\left|\vec{r}_1-\vec{r}_{2j_2}\right|}-\sum_{j_1=1}^{Z_1}\frac{Z_2 e^2}{\left|\vec{r}_2-\vec{r}_{1j_1} \right|}, 
\end{equation} 
where $\vec{r}_1=\vec{r}_{10}+\delta \vec{r}_1$ and $\vec{r}_2=\vec{r}_{20}+\delta \vec{r}_2$ --- vectors determining the position of atomic nuclei; $\vec{r}_{1j_1}=\vec{r}_1+\delta \vec{r}_{j_1}$ and $\vec{r}_{2j_2}=\vec{r}_2+\delta \vec{r}_{j_2}$ --- vectors determining the position of the $ j_1$-th electron of the first atom and $j_2$-th electron of the second atom.

\par Fluctuations in the potential interaction energy \eqref{eq1} are caused by quantum fluctuations experienced by atomic electrons and nuclei. Averaging over the quantum fluctuations of the location of atomic electrons will be carried out using the method \cite{Bethe:1964:IQM}, which Bethe used to calculate the atomic form factor, and averaging over the quantum fluctuations of atomic nuclei will be performed as the Debye-–Waller factor is introduced. We average \eqref{eq1} over the squares of the absolute values of the wave functions of electrons and atomic nuclei. The corresponding means will be denoted $\left\langle ...\right\rangle _{e1} $, $\left\langle ...\right\rangle _{e2} $, $\left\langle ...\right\rangle _{n1} $, $\left\langle ...\right\rangle _{n2} $.

We expand the potential interaction energy \eqref{eq1} in the Fourier integral
\begin{equation} \label{eq2} 
\begin{array}{c} 
{U=\Large \displaystyle{\int\frac{d^3\vec{k}}{(2\pi)^{3}}  \left(\frac{4\pi Z_1 Z_2 e^2}{k^2} \exp{\left[i\vec{k}(\vec{r}_1-\vec{r}_2)\right]}+
\frac{4\pi e^2}{k^2} \sum\limits_{j_1=1}^{Z_1}\sum\limits_{j_2=1}^{Z_2}\exp{\left[i\vec{k}(\vec{r}_{1j_1}-\vec{r}_{2j_2})\right]}-\right.}} \\ 
\Large \displaystyle{{\left. -\frac{4\pi Z_2 e^2}{k^2} \sum\limits_{j_1=1}^{Z_1}\exp{\left[i\vec{k}(\vec{r}_2-\vec{r}_{1j_1})\right]}-\frac{4\pi Z_1 e^2}{k^2} \sum\limits_{j_2=1}^{Z_2}\exp{\left[i\vec{k}(\vec{r}_1-\vec{r}_{2j_2})\right]}\right).}}
\end{array} 
\end{equation} 

\par We average \eqref{eq2} over the squared modulus of the wave function of the electrons of the first atom
\begin{equation} \label{eq3} 
\begin{array}{c} 
{\left\langle U\right\rangle _{e1} =\Large \displaystyle\int \frac{d^3\vec{k}}{(2\pi)^3} \left(\frac{4\pi Z_1 Z_2 e^2}{k^2} \exp\left[i\vec{k}(\vec{r}_1-\vec{r}_2)\right]+\right.  \frac{4\pi e^2}{k^2} \left\langle\sum\limits_{j_1=1}^{Z_1}\sum\limits_{j_2=1}^{Z_2}\exp\left[i\vec{k}(\vec{r}_{1j_1}-\vec{r}_{2j_2})\right]\right\rangle_{e1}-} \\ 
\Large \displaystyle{\left. -\frac{4\pi Z_2 e^2}{k^2} \left\langle \sum\limits_{j_1=1}^{Z_1}\exp\left[i\vec{k}(\vec{r}_2-\vec{r}_{1j_1})\right] \right\rangle_{e1}-\frac{4\pi Z_1 e^2}{k^2}\sum\limits_{j_2=1}^{Z_2}\exp\left[i\vec{k}(\vec{r}_1-\vec{r}_{2j_2})\right] \right);} 
\end{array} 
\end{equation} 
$$\left\langle \sum\limits_{j_1=1}^{Z_1}\sum\limits_{j_2=1}^{Z_2}\exp\left[i\vec{k}(\vec{r}_{1j_1}-\vec{r}_{2j_2})\right] \right\rangle_{e1}=F_1(k)\sum\limits_{j_2=1}^{Z_2}\exp\left[i\vec{k}(\vec{r}_1-\vec{r}_{2j_2})\right];
$$ 
$$\left\langle \sum\limits_{j_1=1}^{Z_1}\exp\left[i\vec{k}(\vec{r}_2-\vec{r}_{1j_1})\right]\right\rangle_{e1}=F_1(k)\exp\left[i\vec{k}(\vec{r}_1-\vec{r}_2)\right],
$$
where $F_1(k)$--atomic form factor; $F_1(0)=Z_1$.

We average $\left\langle U\right\rangle_{e1}$ over the squared modulus of the wave function of the electrons of the second atom
$$
\left\langle U\right\rangle_{e1,e2} =\int \frac{d^3\vec{k}}{(2\pi)^3}\left(\frac{4\pi Z_2 e^2}{k^2} \left(Z_1-F_1(k)\right)\exp\left[i\vec{k}(\vec{r}_1-\vec{r}_2)\right]-\frac{4\pi e^2}{k^2} \left(Z_1-F_1(k)\right)\left\langle \sum\limits_{j_2=1}^{Z_2}\exp\left[i\vec{k}(\vec{r}_1-\vec{r}_{2j_2})\right]\right\rangle_{e2}\right);
$$ 
\begin{equation}
\left\langle \sum\limits_{j_2=1}^{Z_2}\exp\left[i\vec{k}(\vec{r}_1-\vec{r}_{2j_2})\right]\right\rangle_{e2}=F_2(k)\exp\left[i\vec{k}(\vec{r}_1 -\vec{r}_2)\right].
\end{equation} 

\par Let us average the potential energy of interaction of two atoms over the squares of the moduli of the wave functions of atomic nuclei, which slightly deviate from their equilibrium position due to quantum fluctuations
\begin{equation}
\left\langle U\right\rangle_{e1,e2,n1,n2} =\int\frac{d^3\vec{k}}{(2\pi)^3}\frac{4\pi e^2}{k^2}\left[Z_1-F_1(k)\right]\left[Z_2-F_2(k)\right]\exp\left[-\frac{k^2}{2}(\sigma_1^2+\sigma_2^2)\right]\exp\left[i\vec{k}(\vec{r}_{10}-\vec{r}_{20})\right], 
\end{equation} 
where $U(k)=\frac{4\pi e^2}{k^2} \left[Z_1-F_1(k)\right]\left[Z_2-F_2(k)\right]\exp\left[-\frac{k^2}{2}(\sigma_1^2+\sigma_2^2)\right]$--Fourier component of the potential energy of interaction of two atoms; $\sigma_1^2$ and $\sigma_2^2$ -- average squares of the amplitude of quantum fluctuations of atomic nuclei per one degree of freedom; $\vec{r}=\left|\vec{r}_{10}-\vec{r}_{20}\right|$ -- distance between atoms.

\par Similar to how this is done in kinetic theory \cite{pitaevskii2012physical}, we add to the expression for the Fourier component of the potential energy of interaction of two atoms a factor $\left(1-F(k)/Z\right)$, with which we will take into account the Pauli principle. The quantity $F(k)/Z$ is the Fourier component of the distribution density of atomic electrons, which is normalized to unity. As a result, we get
\begin{equation}
U(k)=\frac{4\pi Z_1 Z_2 e^2}{k^2} \left[1-\frac{F_1(k)}{Z_1}\right]^2 \left[1-\frac{F_2(k)}{Z_2}\right]^2\exp\left[-\frac{k^2}{2}(\sigma_1^2+\sigma_2^2)\right].  
\end{equation} 

\par Atomic form factors were calculated using the wave functions of a hydrogen-like atom
\begin{equation}
F_i(k)=\frac{16n_{i1}}{(4+a_i^2 k^2)^2}+\frac{n_{i2}(1-3a_i^2 k^2+2a_i^4 k^4)}{(a_i^2 k^2+1)^4}+\frac{n_{i3}(1-5a_i^2 k^2)}{(a_i^2 k^2+1)^4} +\frac{n_{i4}}{(1+a_i^2 k^2)^3} ,  
\end{equation} 
where $n_{ij}$ -- electron distribution over electron shells $1s,2s,2p_0,2p_{\pm 1} $ for the $i$-th atom; subscript in $2p_0$ and $2p_{\pm 1}$--states indicates the value that the magnetic quantum number takes $m=0$ or $m=\pm 1$; $a_{i} =a_0/Z_i \approx 0.529{\mathop{{\rm A} }\limits^{\circ }} / Z_{i}$ --screening length $i$-th of a hydrogen atom. 

\par In $1s,2s,2p_0$ -- states, there can be from zero to two electrons ($0\le n_{ij}\le 2$ at $1\le j \le 3 $), and in $2p_{\pm 1}$ -- state from zero to four electrons ($0\le n_{i4}\le 4$).

\par We will follow the principle that a closed physical system seeks to achieve a state with the least energy. It can be shown that the potential interaction energy of two beryllium atoms reaches a minimum when the electrons are distributed over the electron shells as follows

$$
(n_{ij} )=\left(
\begin{array}{cccc} 
{0} & 2 & 2 & 0 \\ 
{0} & 2 & 2 & 0 
\end{array}
\right),\textrm{~that's~~} 
\left(
\begin{array}{cccc}
0 & 2s^2 & 2p_0^2 & 0 \\
0 & 2s^2 & 2p_0^2 & 0 
\end{array}\right).$$

\par A graph of the potential interaction energy of two beryllium atoms as a function of the distance between them is shown in Fig.1. A comparison of the calculation results with experiment \cite{Merritt1548} is presented in Fig.\ref{fig2}. Satisfactory agreement with the experiment is observed along the depth of the potential well, but its position and width differ from the experimental results \cite{Merritt1548}. If the value of the screening length is reduced, the position and depth of the potential well will be close to the experimental results, and the width of the well will differ from the experimental values. Further convergence of the two beryllium atoms leads to a new potential minimum (Fig.\ref{fig1a}), which could not be found in the literature. It can be seen that this new quasistationary state will not be stable, since the charge state of the beryllium dimer can change due to the tunneling effect of atomic electrons through the barrier (Fig.\ref{fig1b}). The question remains open about the possibility of observing in experiment the effects associated with this new quasistationary state. The calculation results with $\sigma>0$ will be published later.
\\
\\
\section*{The Acknowledgements}
\par The reported study was funded by RFBR, project number 20-07-00236 a.

\section*{Conflict of interest}
\par The authors declare that they have no conflict of interest.

\bibliographystyle{unsrturl}%unsrt
\bibliography{refs}

\newpage
\section*{\center List of figures}
\par \textbf{Fig. 1}. Potential interaction energy of two beryllium atoms depending on the distance between them: a)$r/a\in[0;11]$, b)$r/a\in[10;20]$, c)$r/a\in[18;32]$. Calculation result for $a_i=0.529/Z_i=0.1323{\mathop{{\rm A} }\limits^{\circ }}$(dashed line) and $a_i=0.117{\mathop{{\rm A} }\limits^{\circ }}$(dash-dot line). In all calculations $\sigma=0$.
\par \textbf{Fig. 2}. Comparison of experiment \cite{Merritt1548} (solid line) and calculation results for $a_i=0.529/Z_i=0.1323\AngM$(dashed line) and $a_i=0.117{\mathop{{\rm A} }\limits^{\circ }}$(dash-dot line). In all calculations $\sigma=0$.

\newpage
\begin{figure}[!b]
  \begin{center}
    \includegraphics[width=\textwidth]{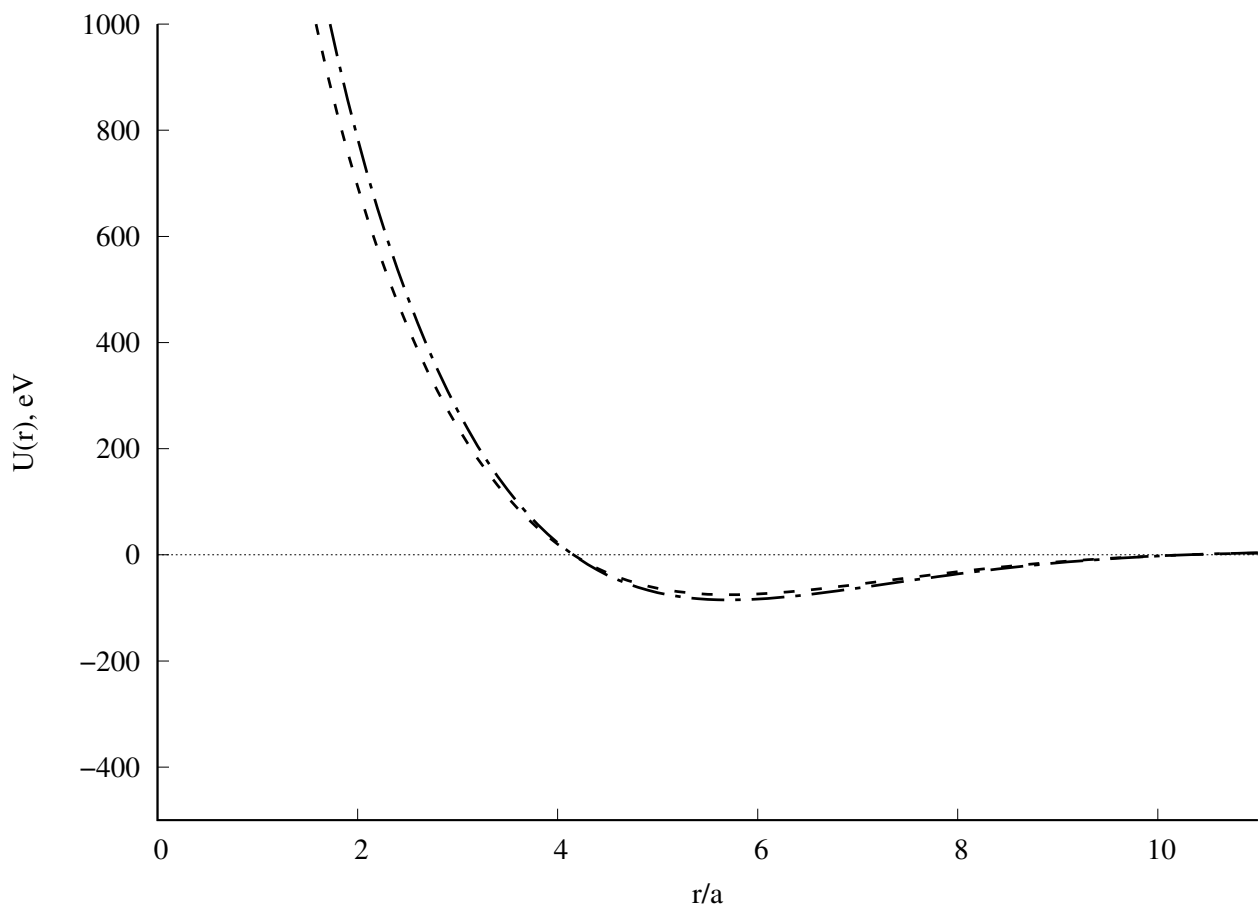}
  \end{center}

  \caption{\small Potential interaction energy of two beryllium atoms depending on the distance between them: $r/a\in[0;11]$. Calculation result for $a_i=0.529/Z_i=0.1323{\mathop{{\rm A} }\limits^{\circ }}$(dashed line) and $a_i=0.117{\mathop{{\rm A} }\limits^{\circ }}$(dash-dot line). In all calculations $\sigma=0$.}
  \label{fig1a}
  \addtocounter{subfig}{1}
  \setcounter{figure}{0}
\end{figure}

\newpage
\begin{figure}[!b]
  \begin{center}
  	\includegraphics[width=\textwidth]{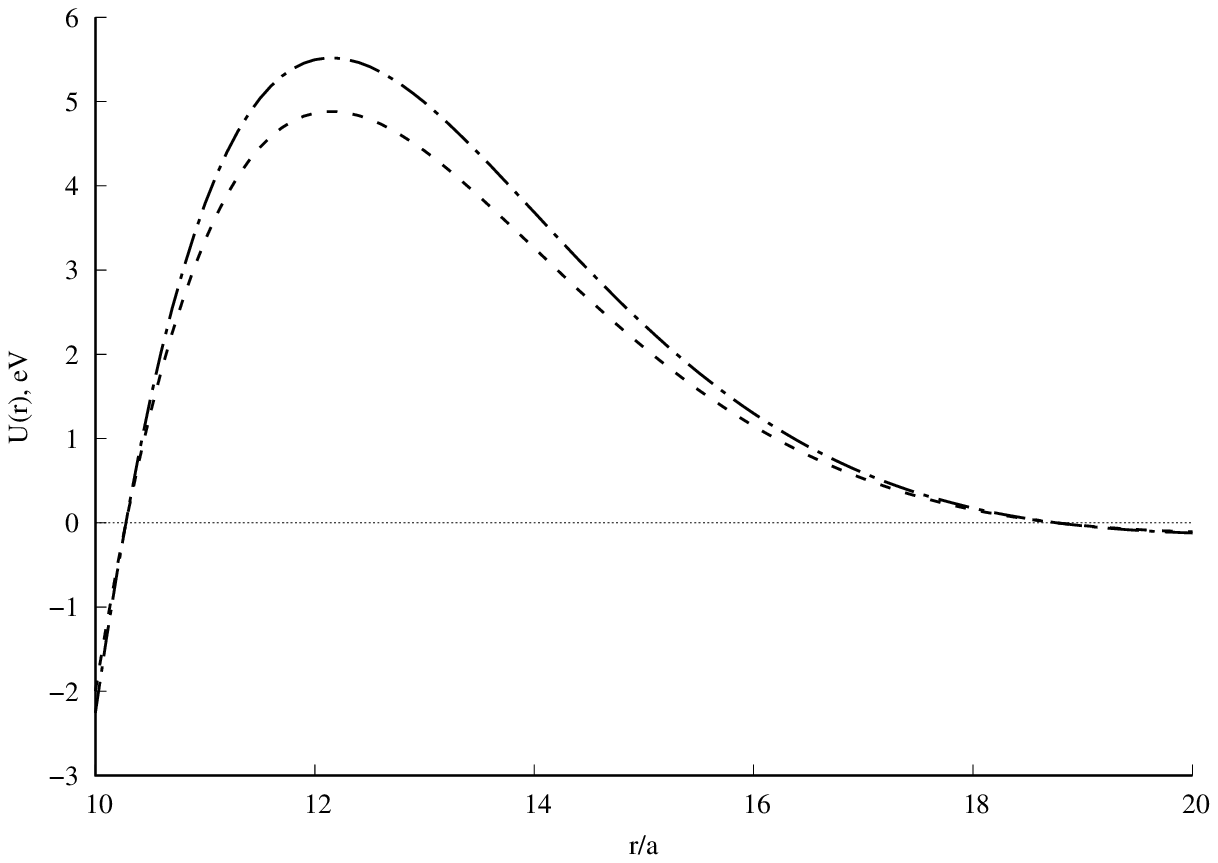}
  \end{center}

  \caption{\small Potential interaction energy of two beryllium atoms depending on the distance between them: $r/a\in[10;20]$. Calculation result for $a_i=0.529/Z_i=0.1323{\mathop{{\rm A} }\limits^{\circ }}$(dashed line) and $a_i=0.117{\mathop{{\rm A} }\limits^{\circ }}$(dash-dot line). In all calculations $\sigma=0$.}
  \label{fig1b}
  \addtocounter{subfig}{1}
  \setcounter{figure}{0}
\end{figure}

\newpage
\begin{figure}[!b]
  \begin{center}
  	\includegraphics[width=\textwidth]{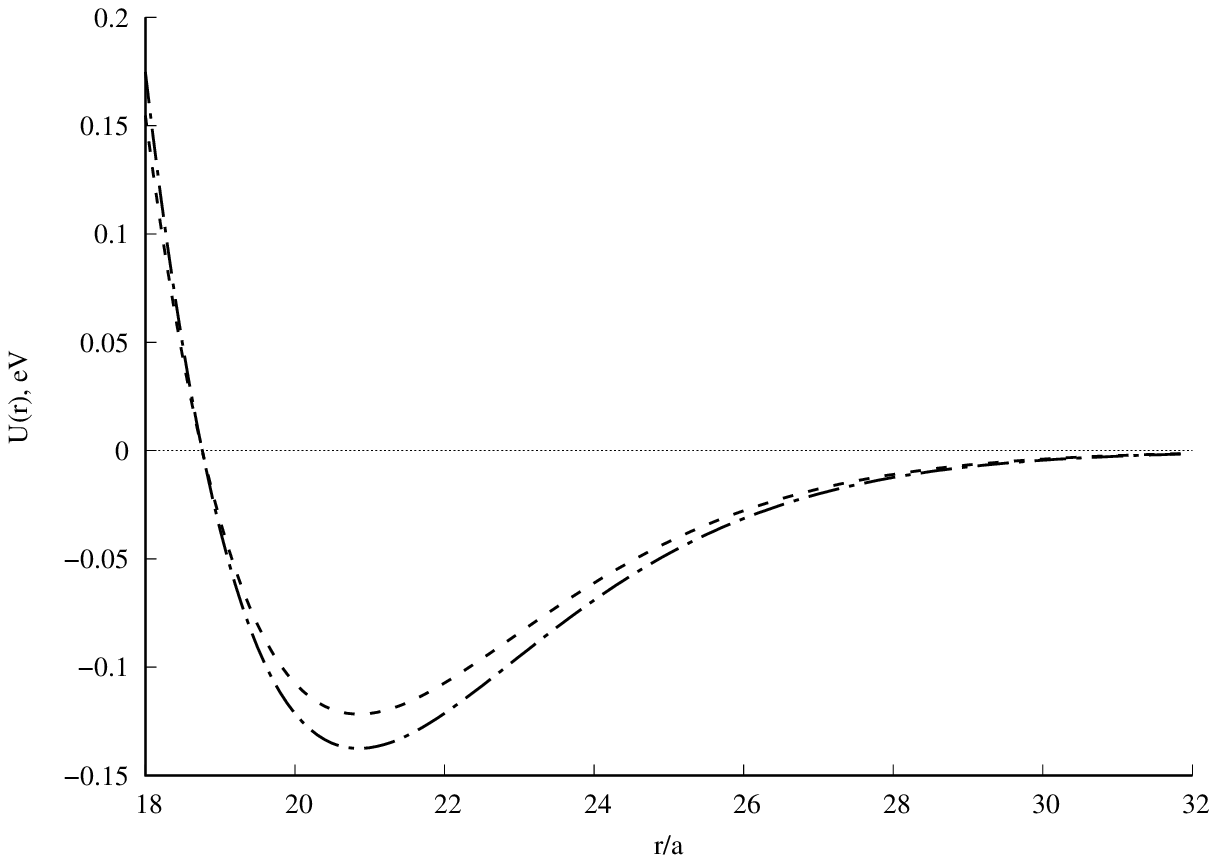}
  \end{center}

  \caption{\small Potential interaction energy of two beryllium atoms depending on the distance between them: $r/a\in[18;32]$. Calculation result for $a_i=0.529/Z_i=0.1323{\mathop{{\rm A} }\limits^{\circ }}$(dashed line) and $a_i=0.117{\mathop{{\rm A} }\limits^{\circ }}$(dash-dot line). In all calculations $\sigma=0$.}
  \label{fig1c}
  \setcounter{subfig}{0}
\end{figure}

\newpage
\begin{figure}[!b]
  \begin{center}
  	\includegraphics[width=\textwidth]{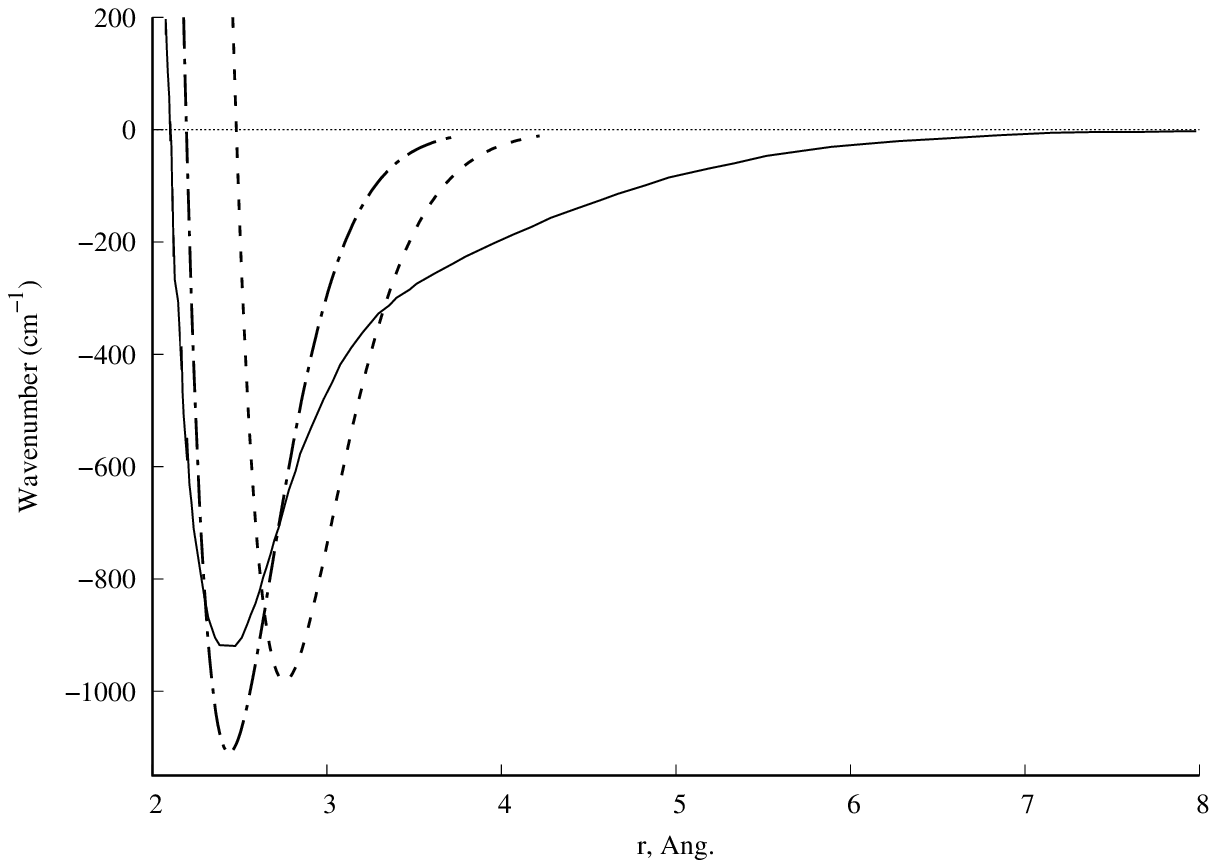}
  \end{center}

  \caption{\small Comparison of experiment \cite{Merritt1548} (solid line) and calculation results for $a_i=0.529/Z_i=0.1323\AngM$(dashed line) and $a_i=0.117{\mathop{{\rm A} }\limits^{\circ }}$(dash-dot line). In all calculations $\sigma=0$.}
  \label{fig2}
\end{figure}

\end{document}